# New Astronomy Reviews Special Issue: History of *Kepler*'s Major Exoplanet "Firsts"

Jack J. Lissauer
Space Science & Astrobiology Division, MS 245-3
NASA Ames Research Center
Moffett Field, CA 94035, USA

&

Joann Eisberg
School of Mathematics & Science, ZH 115
Chaffey College
5885 Haven Avenue
Rancho Cucamonga, CA 91737, USA

NASA's *Kepler* Mission revolutionized exoplanet science in the early part of the 2010's. Looking back from the perspective of the end of that decade, *Kepler* appears to have burst upon the scene ready for battle, like Athena springing forth, fully formed, from the head of Zeus. The story was not so simple. *Kepler*'s first major exoplanet discoveries were not announced until more than a year had passed since the spacecraft started collecting scientific data, and by that time many exoplanet scientists not working on the project had become frustrated with the lack of results coming from the *Kepler* project. But an immense amount of work was required to develop the tools and conceptual framework needed to harvest the abundant field of data that the spacecraft was producing. This issue contains articles describing some of the research efforts, most of which took place behind the scenes, that led to the announcements/publication of several of *Kepler*'s major exoplanet "firsts", written by the scientists who led the landmark discovery papers. These papers were all published between 1½ and 4½ years after launch, and many were accompanied by press events designed to share the findings with as large an audience as possible, especially U.S. taxpayers, who had provided funding for the mission.

*Kepler* was a watershed in the study of extrasolar planets. The majority of exoplanets currently known, including the vast majority of small exoplanets, are *Kepler* discoveries. Although *Kepler*'s primary mission was to conduct a statistical census of the exoplanet population, along the way it found dozens of especially novel and interesting types of exoplanets and exoplanetary systems, many with characteristics scarcely imagined when *Kepler* was launched. *Kepler* also marks a significant development of new methods for exoplanet data analysis. *Kepler's* profusion of planet candidates overwhelmed the resources available for ground-based Radial Velocity observations to confirm planetary status, and other strategies for planet confirmation were urgently needed.

To mark the tenth anniversary of the commencement of *Kepler* science operations, *New Astronomy Reviews* is publishing this Virtual Special Issue containing articles that describe the stories behind several of *Kepler*'s seminal exoplanet discoveries and methodological innovations. The styles of the articles are as diverse as the planets whose discoveries are described therein, with many focusing on the personal histories that led researchers to their discoveries and others focusing more on updates to our knowledge of these watershed planets and systems. All articles have been refereed by experts in the specific topics of the discoveries and reviewed by a historian of astronomy to ensure high quality and relevance to a diverse audience. Specialists and non-specialists alike will be interested by themes including: detailed insight into the practices of modern astrophysical research, the human dynamics of collaboration and competition in this large community of driven and hard-working researchers, the opportunities and constraints of big data sets and the analysis thereof, the tension between the primary mission of *Kepler* to conduct a statistical census and the prestige and newsworthiness of discoveries of individual planets and planetary systems.

*Kepler* delighted researchers and the public with the variety of exoplanets discovered. Seven of the eight manuscripts thus far accepted for publication in this issue focus in whole or in part on landmark discoveries of individual planets or planetary systems.

Kepler-16b provided unequivocal proof of a bound third body in a circumbinary orbit around two stars, and the discovery paper showed how much information about the system could be derived when a planet was observed to transit two stars that themselves composed an eclipsing binary. Doyle's article on the history of its discovery ([arXiv:1908.02838](arXiv:1908.02838)) also reminds us of the tremendous public enthusiasm for *Kepler*'s results, epitomized by the nick-naming of this world after the Star Wars planet "Tatooine".

*Kepler*'s key mission goal was to determine the occurrence rate of earthlike planets, and several articles address the first Earth-sized planets. Discovering distant worlds the size of our own has been a long-held dream of astronomers. Kepler-20e and Kepler-20f finally realized that dream. Torres and Fressin ([arXiv:1905.04309](arXiv:1905.04309)) make the point that the masses of Kepler-20e and Kepler-20f were too small to measure with Doppler techniques, necessitating the development of statistical techniques for validation.

Kepler-62f is the first planet not much larger than the Earth to have been discovered orbiting in its star's habitable zone, and it arguably remains the known exoplanet most likely to be habitable. Despite being identified as a planet candidate in early *Kepler* catalogs and being validated as such in the discovery paper, it was labeled a false positive by the "Robovetter," an algorithm designed to ensure uniformity of criteria of credibility of planet candidates. The article by Borucki, Thompson, Agol and Hedges ([arXiv:1905.05719](arXiv:1905.05719)) argues that planet candidates as high-value as Kepler-62f do indeed merit individual consideration; comprehensive analysis of all available data demonstrate that Kepler-62f is a real planet.

Other planets are remarkable for the extreme locations. The article by Winn, Sanchis-Ojeda and Rappaport ([arXiv:1803.03303](arXiv:1803.03303)) describes Kepler-78 and the Ultra-Short-Period planets, which may help us to understand the formation and orbital evolution of short-period planets, as well as star-planet interactions, atmospheric erosion, and other phenomena arising from strong irradiation and strong tidal forces.

Key among *Kepler's* discoveries have been large numbers of multi-planet systems. Planet-planet interactions in such systems permit precise measurements of system parameters and allow astronomers to probe the density (and hence the composition) of planets. Ragozzine and Holman's manuscript ([arXiv:1905.04426](arXiv:1905.04426)) describes the discovery of two planets slightly smaller than Saturn and near the 2:1 mean orbital motion resonance in the Kepler-9 system. This made Kepler-9 the first system with more than one transiting planet, and these sub-Saturns were also first planets observed to have Transit Timing Variations (TTVs). TTVs are departures of the transits from uniform spacing in time, typically caused by gravitational perturbations from other planets. The article by Agol and Carter ([arXiv:1905.05229](arXiv:1905.05229)) tells of the development of TTV techniques and photodynamical modelling used in the unprecedentedly precise characterization of Kepler-36 and its system. The article by Nesvorny ([arXiv:1905.04262](arXiv:1905.04262)) describes the Kepler-46 system, in which TTVs were first used to detect and precisely characterize a non-transiting a planet, increasing the completeness of our planetary survey.

Finally, the paper by Steffen and Lissauer ([arXiv:1905.04659](arXiv:1905.04659)) addresses not a specific system, but rather two early papers about the discovery and analysis of *Kepler*'s multi-transiting population. The large number of such systems found by *Kepler* implies that flat compact multi-planet systems are common. Their characteristics have revealed much about the dynamical architectures of planetary systems, and they provide useful constraints on models the formation of exoplanetary systems. New statistical techniques were developed to confirm or validate hundreds of candidate multi-planet systems.

We are hoping to add to the above list later this year with additional scheduled manuscripts reviewing the discoveries of *Kepler*'s first rocky planet, Kepler-10 b, the first transiting high-multiplicity system, Kepler-11, which hosts six planets observed to transit their sunlike star, the production of the first two *Kepler* catalogs of planet candidates, and the first detailed estimates of planet occurrence rates from *Kepler* data.

We thank the authors for taking time from their busy schedules to write the articles that appear herein, Dr. Cathie Clarke for handling the editorial duties for the article (and hopefully soon a second article) on which JJL was a co-author, and many referees (the majority of whom chose to be anonymous) for providing thorough and constructive reviews of each article.